\documentclass[journal=ancac3,manuscript=article,layout=twocolumn]{achemso}
\setkeys{acs}{articletitle=true}
\setkeys{acs}{keywords = true}

\usepackage{tabularx}
\usepackage[flushleft]{threeparttable}
\usepackage{chemformula}
\usepackage{lineno}
\usepackage{multirow}
\usepackage{hhline}
\usepackage{amssymb}
\usepackage{color}
\usepackage{amsmath}
\usepackage{relsize}
\usepackage[font=scriptsize,labelfont=bf]{caption}
\usepackage{}
\usepackage{ulem}
\usepackage{colortbl}
\usepackage{threeparttable}

\usepackage[version=3]{mhchem}
\usepackage{booktabs} 
\usepackage[symbol]{footmisc}
\usepackage{caption}
\usepackage{booktabs}

\title{Specific Iron Binding to Natural Sphingomyelin Membrane Induced by Non-Specific Co-Solutes}

\author{Wenjie Wang}
\affiliation{Division of Materials Sciences and Engineering, Ames National Laboratory, U.S. DOE, Ames, Iowa 50011, United States}
\author{Honghu Zhang}
\affiliation{Department of Materials Science and Engineering, Iowa State University, Ames, Iowa 50011, United States}
\author{Binay P. Nayak}
\affiliation{Department of Chemical and Biological Engineering, Iowa State University, Ames, Iowa 50011, United States}
\author{David Vaknin} 
\affiliation{Ames National Laboratory and Department of Physics and Astronomy, Iowa State University, Ames, Iowa 50011,  United States}

\email{vaknin@ameslab.gov}
\normalsize
\begin{document}
\flushbottom
\maketitle
\normalsize
\begin{abstract} 

\noindent\textbf{Hypothesis:}
\noindent Sphingomyelin (SPM), a crucial phospholipid in the myelin sheath, plays a vital role in insulating nerve fibers. We hypothesize that iron ions selectively bind to the phosphatidylcholine (PC) template within the SPM membrane under near-physiological conditions, resulting in disruptions to membrane organization. These interactions could potentially contribute to the degradation of the myelin sheath, thereby playing a role in the development of neurodegenerative diseases.

\noindent\textbf{Experiments:}
\noindent We utilized synchrotron-based X-ray spectroscopy and diffraction techniques to study the interaction of iron ions with a bovine spinal-cord SPM monolayer (ML) at the liquid-vapor interface under physiological conditions. The SPM ML serves as a model system, representing localized patches of lipids within a more complex membrane structure. The experiments assessed iron binding to the SPM membrane both in the presence of salts and with additional evaluation of the effects of various ion species on membrane behavior. Grazing incidence X-ray diffraction was employed to analyze the impact of iron binding on the structural integrity of the SPM membrane.

\noindent\textbf{Findings:}
\noindent Our results demonstrate that iron ions in dilute solution selectively bind to the PC template of the SPM membrane exclusively at near-physiological salt concentrations (e.g., NaCl, KCl, KI, or CaCl$_2$) and is pH-dependent. In-significant binding was detected in the absence of these salts or at near-neutral pH with salts. The surface adsorption of iron ions is correlated with salt concentration, reaching saturation at physiological levels. In contrast, multivalent ions such as La$^{3+}$ and Ca$^{2+}$ do not bind to SPM under similar conditions. Notably, iron binding to the SPM membrane disrupts its in-plane organization, suggesting that these interactions may compromise membrane integrity and contribute to myelin sheath damage associated with neurological disorders.

\end{abstract}

\thispagestyle{empty}

\section{Introduction}
Sphingomyelin (SPM) and di-palmitoyl phosphatidylcholine are key membrane lipids primarily located in the plasma membrane leaflet. They serve as crucial structural components and play a significant role in cell signaling \cite{Barenholz1999, Jao2009}. SPM is particularly abundant in the membranous myelin sheath surrounding nerve cell axons \cite{Ramstedt2002}. Dysfunction in the peripheral nerve and brain tissue associated with SPM may contribute to various neurodegenerative disorders \cite{Yadav2014}.

It has been reported that the breakdown of the myelin sheath is linked to a decline in brain cognitive function \cite{Bartzokis2004}. In addition, elevated levels of iron ions have been observed in brain tissues affected by neurodegenerative diseases \cite{Zecca2004}. Iron (Fe), an essential element, plays a vital role in various cellular processes in the brain, including myelin synthesis \cite{Ndayisaba2019}. Progressive accumulation of iron with age has been identified in damaged areas of the brain afflicted by various diseases \cite{Smith1997, Raven2013, Kell2010}. Monitoring changes in iron levels is speculated to be a marker of disease progression. The role, location, and ionic form of iron in neurodegenerative diseases, such as Alzheimer's and Parkinson's diseases, remain debated \cite{Zecca2004, Ndayisaba2019}. Recently it has been demonstrated that iron accumulation in myelin-related molecular systems can affect cognitive and psychiatric functions, highlighting the role of myelin in neurodegenerative diseases. Increased iron levels in brain tissues are correlated with the severity of multiple sclerosis, underscoring the importance of iron regulation in maintaining neurological health. \cite{heidari2016brain,heidari2016pathological,khattar2021investigation,ektessabi1999distribution}To investigate the spatial location and nature of iron in diseased tissues, synchrotron X-ray techniques have emerged as promising methods \cite{Collingwood2008, Collingwood2014}.

No systematic physicochemical experimental studies have been conducted that directly connect iron and SPM. Recently, using another central phospholipid membrane, 1,2-dipalmitoyl-sn-glycero-3-phosphocholine (DPPC), which shares similar membranous properties with SPM due to their phosphatidylcholine (PC) groups, we uncovered some peculiarities of iron binding to the PC template formed by DPPC at the aqueous/vapor interface, as probed by surface-sensitive X-ray scattering and spectroscopic techniques \cite{Wang2016,wieland2013formation}. In particular, it is found that iron, even in dilute concentrations, has a high affinity for binding to the DPPC monolayer (ML) and disrupts its orderly in-plane structure, but only in the presence of KCl at physiological concentration levels ($\sim$100 mM). Additionally, studies have reported the formation of iron-hydroxide clusters at these interfaces.\cite{Wang2016,wieland2013formation} Similar clusters have also been identified in research on membranes, such as those involving ethylenediaminetetracetic acid-based gemini surfactant monolayers.\cite{sowah2021iron}  SPM and DPPC differ in their hydrophobic hydrocarbon chains and interfacial components, resulting in different biological functions \cite{Barenholz1999, Ramstedt2002, Huang1986}. Therefore, direct investigation of the interactions between iron and SPM and the effect of physiological conditions on these interactions is essential. Understanding the iron-SPM interactions could provide the basis for exploring potential physiological iron pathways in iron-associated neurodegenerative diseases.

To simulate the environment of a nerve cell membrane, we use natural SPM extracted from the bovine spinal cord to create a Langmuir ML. This is depicted in Figure 1, where the hydrophilic polar PC headgroup acts as a template in contact with a dilute aqueous Fe(III) solution. In this study, we utilize the SPM monolayer as a simplified model system to investigate membrane properties. While this system offers an ideal platform for studying lipid behavior, it is important to recognize that biological membranes, such as the myelin sheath, contain additional components, including cholesterol and proteins, contributing to their complex structure and function.\cite{rasband2012myelin} Extensive studies have suggested that cholesterol, in conjunction with SPM, plays a crucial role in forming membrane rafts.\cite{lingwood2010lipid,simons1997functional} By focusing on the SPM monolayer, we aim to explore fundamental aspects of lipid organization while acknowledging that real biological systems involve additional factors. This model system is particularly valuable for examining the interactions between iron and SPM, which have significant implications for understanding membrane structure and function. The zwitterionic nature of the PC headgroup, with its combination of charge-neutral and dipolar characteristics, plays a pivotal role in mediating the association between Fe(III) and SPM. To investigate these interactions in detail, we employ surface-sensitive X-ray diffraction and spectroscopic techniques to qualitatively and quantitatively determine the specific binding of ionic species to the SPM. These techniques have been extensively used to study the structures of Langmuir monolayers and the associated molecules and ions that accumulate at their head group regions \cite{Nielsen2011, Vaknin2001a, Tolan1999, Pershan2012}. 

\begin{figure}[ht]
\centering
\begin{tabular}{cc}
\includegraphics[width=0.5 \textwidth]{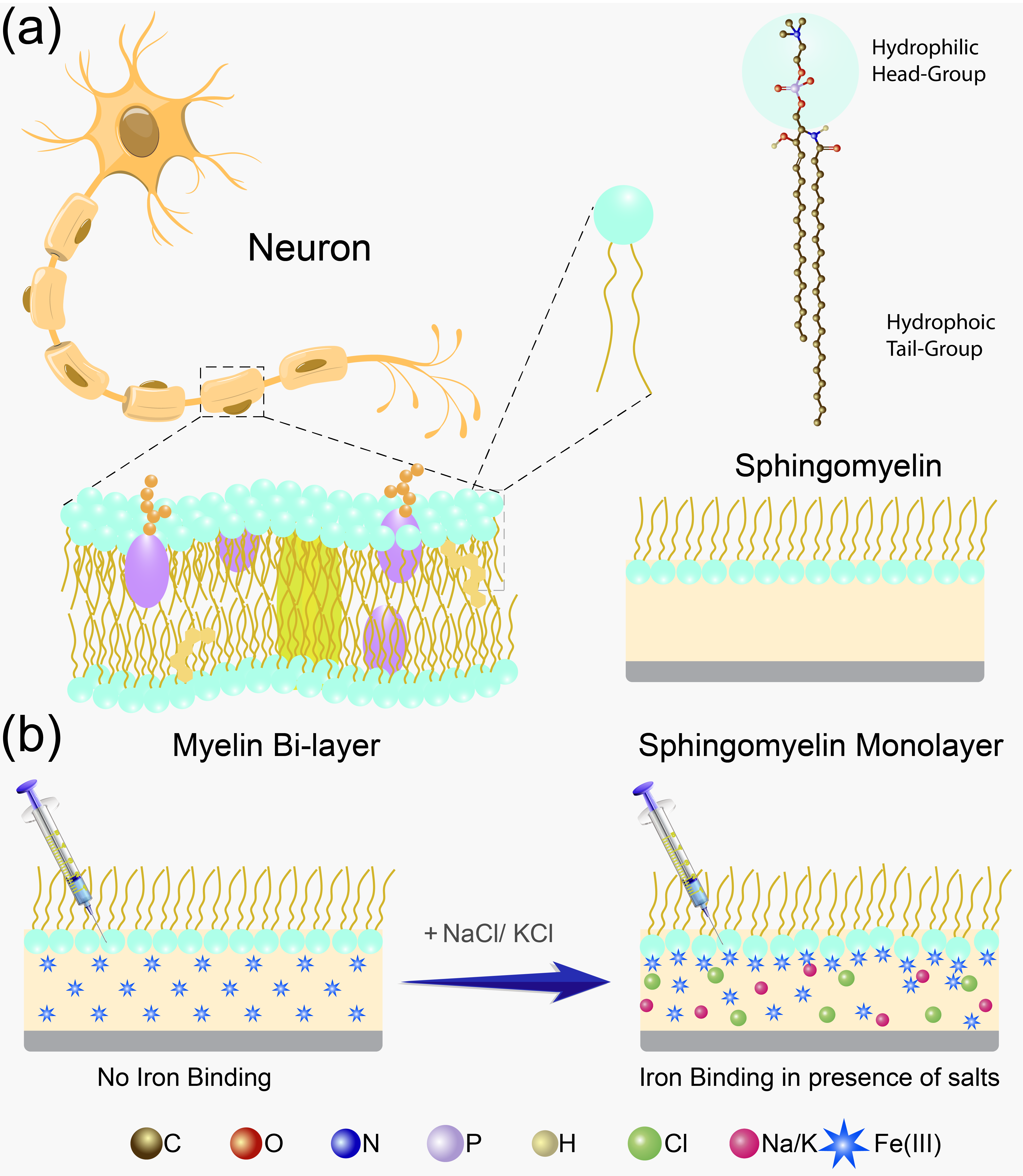}
\end{tabular}
\caption{{\bf Schematics of Fe(III) binding to a sphingomyelin monolayer.} (a) Illustrations depict a neuron wrapped with myelin sheaths, a myelin bilayer membrane, and the molecular structure of a typical sphingomyelin lipid. (b) Depiction of the experimental setup. A sphingomyelin monolayer is formed at the aqueous/vapor interface for surface-sensitive X-ray diffraction and spectroscopic measurements. Fe(III) binding is facilitated by adding salts (e.g., NaCl, KCl) approaching physiological concentrations.}
\label{fig:schematics}
\end{figure}

We systematically explore the binding of aqueous Fe(III) to SPM, a process regulated by pH and the concentrations of other salts (co-solutes). The presence of a sufficient quantity of co-solutes in the aqueous solutions is crucial for inducing surface binding of iron to the SPM monolayer, as depicted in Figure~\ref{fig:schematics} (b). Simple salts, such as KCl and NaCl, constitute indispensable components of buffer solutions for biological systems \cite{Graber2015, Wang2016}. Our findings highlight the essential role of these co-solutes, demonstrating that their presence or absence can dynamically “switch on" and “switch off" surface iron binding to PC templates.

\section{Results and Discussion}

\begin{figure*}[ht]
\centering
\includegraphics[width=0.95 \textwidth]{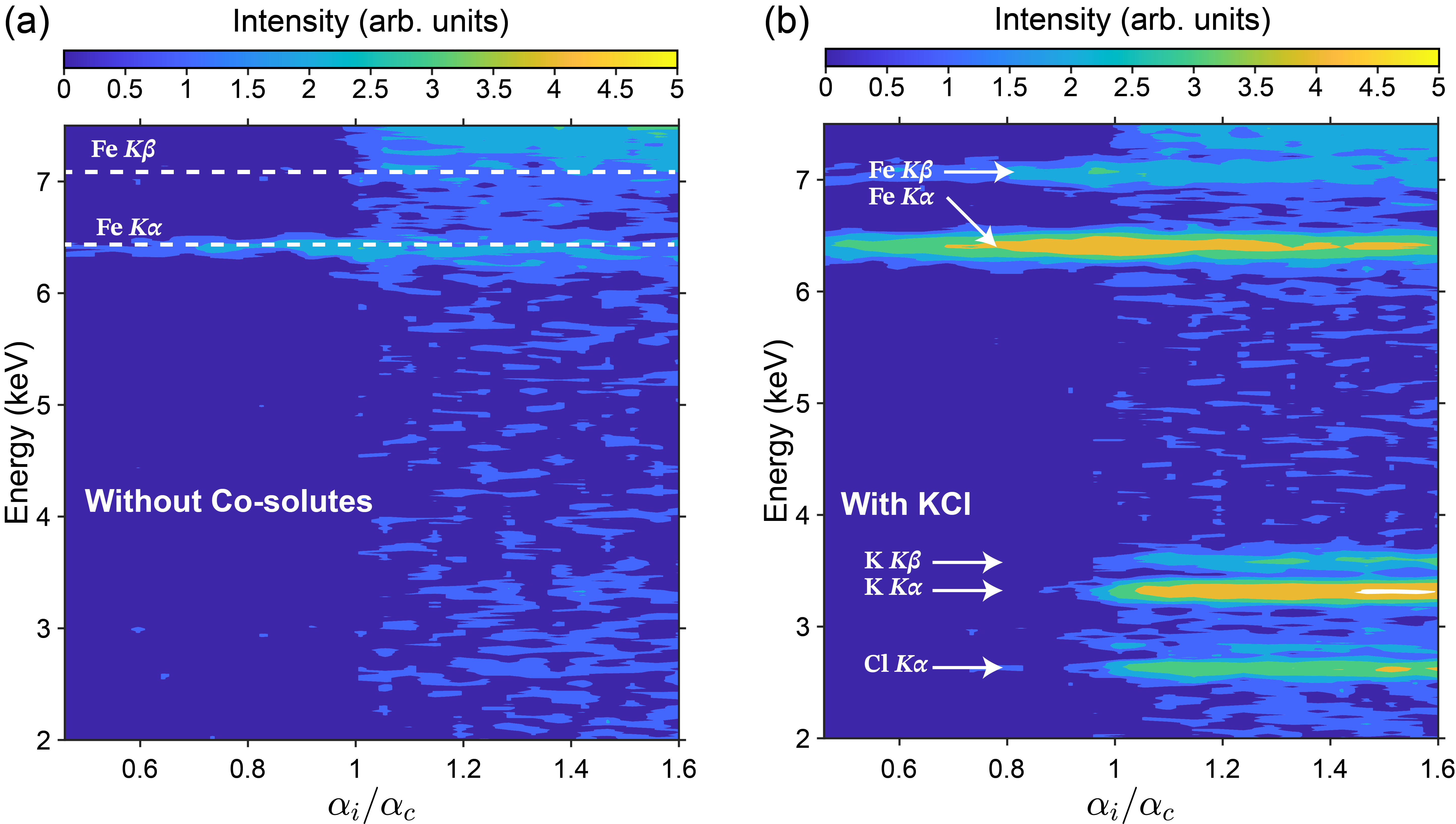}
\caption{ \textbf{Fe(III) binding to SPM monolayer induced by KCl approaching physiological concentration.}
Two-dimensional fluorescence intensity contour plots as a function of emission line fluorescent X-ray energies and the normalized X-ray incident angle ($\alpha_{i} / \alpha_{c}$) for the SPM monolayer over an \ce{FeCl3} subphase (40 $\mu$M, pH 3.0). Panel (a) depicts the absence of any co-solute, showing minimal Fe binding to the SPM. In contrast, panel (b) includes co-solute KCl at approximately 45 mM, revealing strong Fe $K\alpha$ ($\sim 6.4$ keV) and $K\beta$ ($\sim 7.1$ keV) signals near the critical angle. The contour plots partially display the much more intense primary beam (8.05 keV) at the top. Characteristic emission lines for chlorine $K\alpha$ ($\approx 2.6$ keV) and $K\beta$ ($\approx 2.8$ keV), and potassium $K\alpha$ ($\approx 3.3$ keV) and $K\beta$ ($\approx 3.6$ keV), are only observed above the critical angle, with no evidence of bound K or Cl to the SPM monolayer. 
} 
\label{fig:fluores_Ames_2D}
\end{figure*}

\subsection*{Effect of co-solute on iron binding to SPM}

Using the near-total reflection X-ray fluorescence (NTXRF) method, we investigate the specific binding of Fe(III) from dilute solutions to the SPM template. To promote this binding, we introduce various co-solutes (e.g., NaCl, KCl, etc.) at physiological concentrations. Our results demonstrate that iron-binding occurs exclusively in the presence of these added co-solutes. NTXRF, being an element-specific probe, offers unique surface sensitivity, making it ideal for detecting and analyzing this binding phenomenon.\cite{vaknin2002x,bu2020synchrotron} Figure~\ref{fig:fluores_Ames_2D} presents color contour maps of fluorescence intensity as a function of X-ray emission lines versus the X-ray incident angle $\alpha_{i}$ (refer to the Methods Section for more details). Briefly, the fluorescent signal as a function of incident beam angle $\alpha_{i}$ for $\alpha_{i} < \alpha_{c}$ ($\alpha_{c}$ being the critical angle for total reflection) arises from an evanescent wave with a penetration depth of less than $\sim 100$ Å interacting with fluorescent elements, primarily probing specific signals from ion enrichment at the aqueous surface. For ions binding to the template, the intensity increases with $\alpha_{i}$, reaching a maximum at $\alpha_{c}$. The signal below $\alpha_{c}$ is negligible for elements in the solution that do not enrich the surface, but their signal becomes more pronounced above the critical angle $\alpha_{c}$.

Our results indicate that Fe in solution does not readily bind to the SPM template. Figure \ref{fig:fluores_Ames_2D} (a) presents NTXRF data of the SPM monolayer over an iron solution (40 $\mu$M \ce{FeCl3}, pH 3.0) in the absence of co-solutes,  based on repeated measurements taken consecutively over the course of a day. This is evident from the lack of apparent emission lines from Fe in Figure \ref{fig:fluores_Ames_2D} (a). The weak signal observed near the iron \( K\alpha \) emission line (\(\sim 6.4\) keV) is likely due to a trace amount of iron near surface and due to an escape peak, a typical artifact in silicon-based energy dispersive detectors.  It should be noted that at such low iron bulk concentrations, in the micromolar range, the iron emission signal is almost not detectable either at the surface or in the bulk. This demonstrates that Fe does not readily bind to the SPM template in the absence of co-solutes.

Upon injecting KCl from a stock solution, resulting in a final KCl concentration of approximately 45 mM, Fe binds to the SPM template. Figure \ref{fig:fluores_Ames_2D} (b) shows the enhanced Fe $K\alpha$ and $K\beta$ signals near the critical incident angle, $\alpha_{c}$, for total reflection. This demonstrates that Fe(III) ions or their complexes accumulate significantly at the SPM templates only in the presence of KCl. Notably, the signals for K and Cl, although present above $\alpha_c$, are strikingly absent below it. This indicates that the presence of these ions is confined to the bulk, with no detectable enrichment at the interface.

\begin{figure}[ht]
\centering
\includegraphics[width=0.4\textwidth]{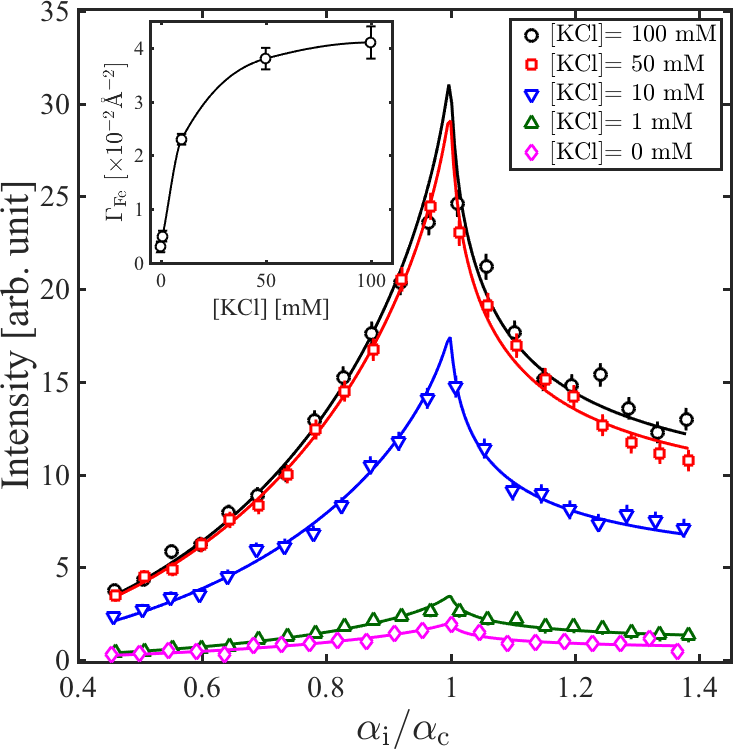}
\caption{\textbf{Fe(III) binding to SPM monolayer scales with KCl concentration.}
Fluorescence intensity integrated over iron $K\alpha$ characteristic emission lines ($6.4 \pm 0.4$ keV) as a function of the normalized X-ray incident angle $\alpha_{i}$ for SPM monolayer spread over \ce{FeCl3} solutions (40 $\mu$M, pH 3.0) containing KCl concentrations of 0, 1, 10, 50, and 100 mM. Scaled by peak intensity, iron binding increases gradually with KCl concentration and saturates at a physiological level of 100 mM. The solid lines are best-fit profiles (see model details in Ref.~[\!\!\citenum{Wang2011b}]). Quantitative iron surface adsorption, $\Gamma_{\rm Fe}$, is obtained by scaling to a pure \ce{FeCl3} bulk solution of known concentration (4 mM, pH 2.0) \cite{Wang2011a, Wang2011b, Wang2016}. The inset displays the iron surface density ($\Gamma_{\rm Fe}$) versus [KCl]. At saturation, $\Gamma_{\rm Fe}$ multiplied by the molecular area of SPM corresponds to approximately 1.6 Fe per SPM molecule.
}
\label{fig:flores_Qz_Ames}
\end{figure}

For a fixed \ce{FeCl3} bulk concentration and a fixed bulk pH level, the quantity of surface-bound Fe in the SPM scales with the concentration of KCl, [KCl]. Figure \ref{fig:flores_Qz_Ames} shows the intensity of Fe $K\alpha$ emission line as a function of $\alpha_{i}$ at various [KCl] as indicated. Analysis of the angular ($\alpha_{i}$)-dependent fluorescent intensity profiles for iron $K\alpha$ characteristic radiation follows the method described in Ref.~[\!\!\citenum{Wang2011b,Wang2011a}]
. By normalizing these profiles with those of a bare surface of Fe solution with known and higher concentration, we determine a surface adsorption of Fe, $\Gamma_{\rm Fe}$. As shown in the inset of Figure \ref{fig:flores_Qz_Ames},  $\Gamma_{\rm Fe}$ increases and saturates at approximately [KCl]=100 mM. We find that the binding of Fe scales with the concentration of KCl, saturating at physiological levels of approximately 100 mM. We note that reversing the sample preparation protocol, namely, spreading SPM directly on solutions containing 40 $\mu$M \ce{FeCl3} mixed with a similar amount of KCl, also results in Fe readily binding to the SPM. This streamlined protocol has henceforth been used to accelerate the measurements.

\begin{figure*}[ht]
\centering
\includegraphics[width=0.95\textwidth]{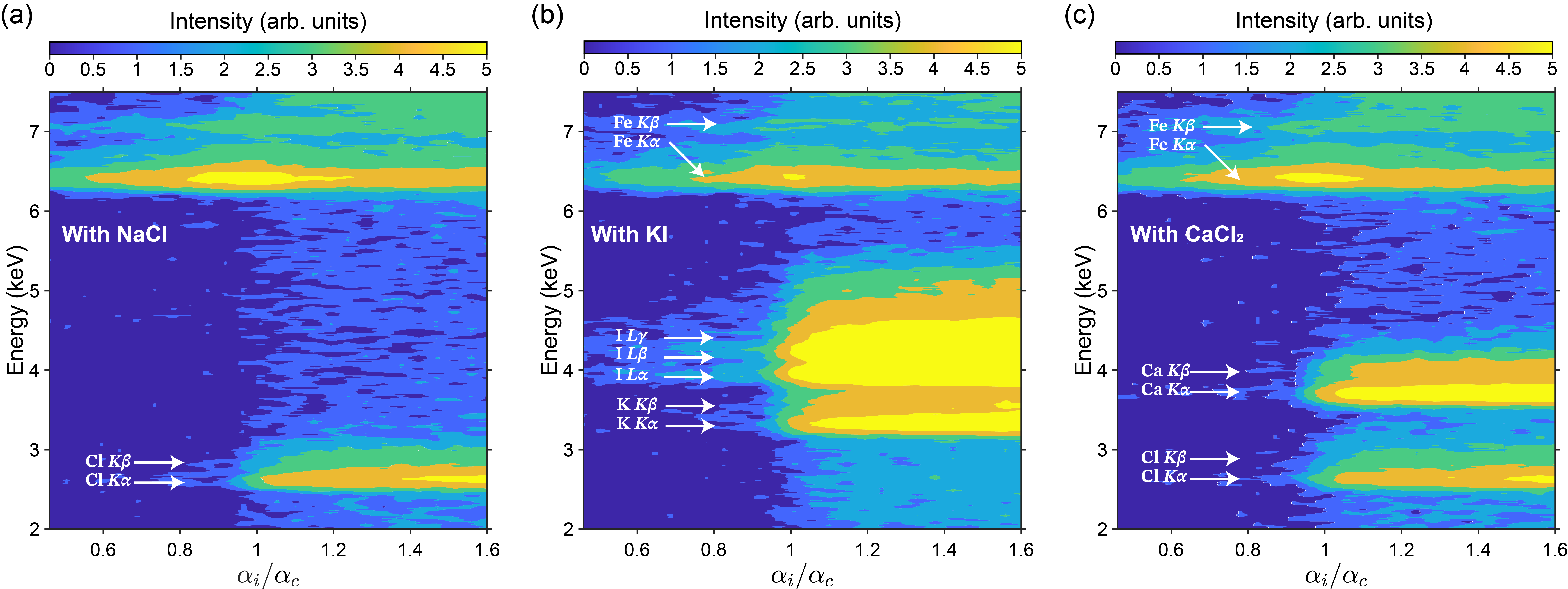} 
\caption{\textbf{Fe(III) binding to SPM can be induced by various salts.}
Two-dimensional fluorescence intensity contour plots as a function of emission line fluorescent X-ray energies and the normalized X-ray incident angle ($\alpha_{i} / \alpha_{c}$) for the SPM monolayer over an \ce{FeCl3} subphase (40 $\mu$M, pH 3.0). In the presence of (a) NaCl 100mM, (b) KI 100 mM, and (c) \ce{CaCl2} 50mM, strong Fe $K\alpha$ ($\sim 6.4$ keV) and $K\beta$ ($\sim 7.1$ keV) signals are observed both below and above the critical angle, indicating strong Fe binding to the SPM monolayer. The contour plots display the tails of the intense primary beam at $\sim$ 8.05 keV. Characteristic emission lines for chlorine $K\alpha$ ($\approx 2.6$ keV) and $K\beta$ ($\approx 2.8$ keV), calcium $K\alpha$ ($\approx 3.7$ keV) and $K\beta$ ($\approx 4.0$ keV), potassium $K\alpha$ ($\approx 3.3$ keV) and $K\beta$ ($\approx 3.6$ keV), and iodine $L\alpha$ ($\approx 3.9$ keV), $L\beta$ ($\approx 4.2$ keV), and $L\gamma$ ($\approx 4.5$ keV) are observed. These lines are only detected above the critical angle, confirming there are no co-solute ions binding to the SPM monolayer.
}
\label{fig:all_salts}
\end{figure*}

The choice of KCl to induce Fe binding is not unique; Fe binding can be induced by various other salts. Figures \ref{fig:all_salts} (a-c) present NTXRF data for the SPM monolayer on dilute \ce{FeCl3} (40 $\mu$M) solutions in the presence of NaCl, KI, and \ce{CaCl2}, each with similar ionic strengths. As mentioned above, and to expedite the process, the SPM films are spread and compressed on mixed Fe/salt solutions. Figures \ref{fig:all_salts} (a-c) show strong Fe $K\alpha$ and $K\beta$ signals below $\alpha_{c}$, indicating strong Fe binding to SPM. Notably, the surface fluorescence signals below $\alpha_{c}$ ($\alpha_{i}/\alpha_{c}<1$) show negligible co-solute ions (i.e., no signals from \ce{K+}, \ce{Ca^2+}, \ce{Cl-}, and \ce{I-}), indicating their absence at the ML/solution interface. Based on these findings, it is concluded that although the presence of other cation and anion species at sufficient concentrations promotes iron-SPM binding, they do not bind to the SPM.

\begin{figure}[ht]
\centering
\includegraphics[width=0.45\textwidth]{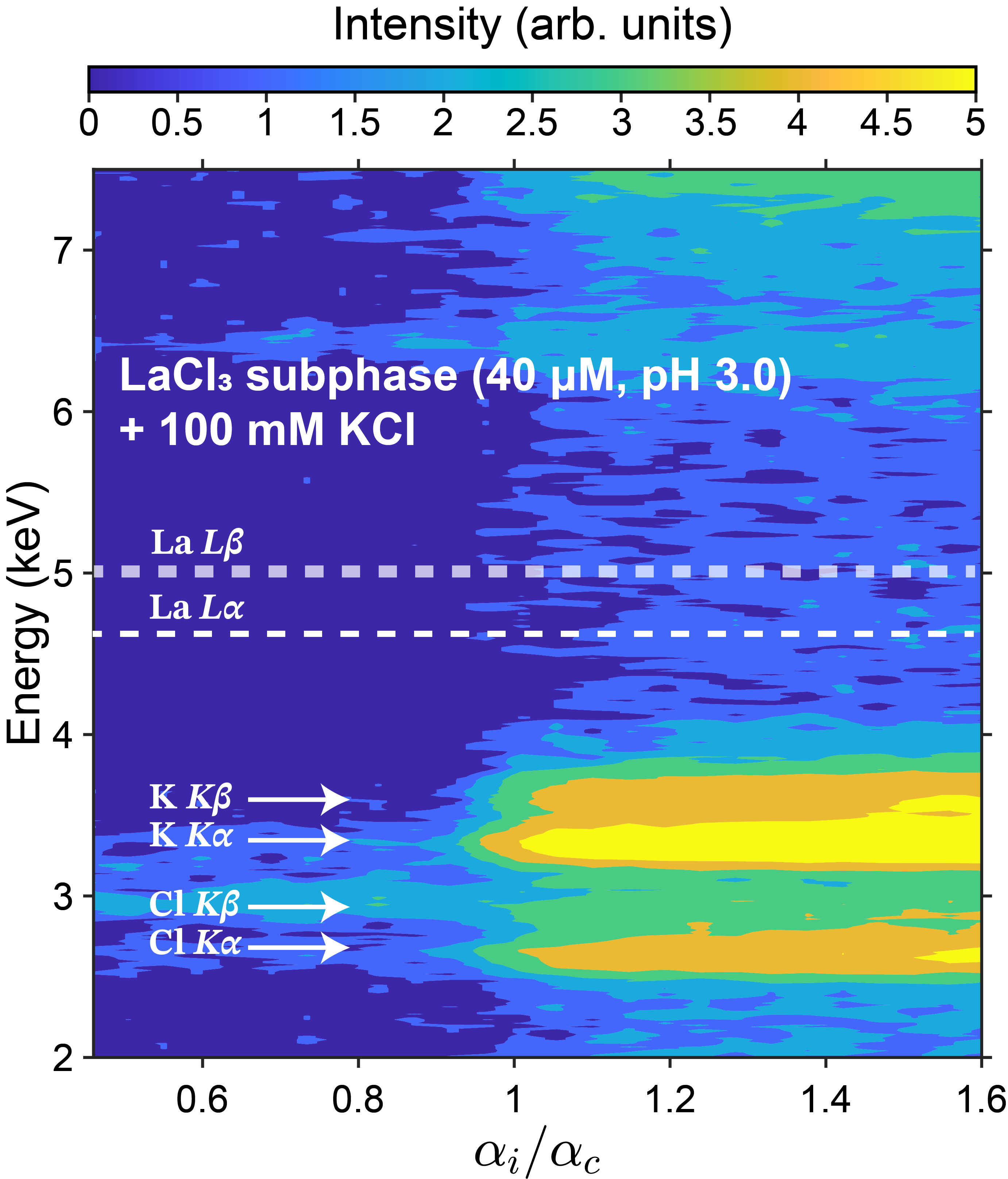}
\caption{\textbf{Non-electrostatic nature of Fe (III) binding to SPM.}
Two-dimensional fluorescence intensity contour plots as a function of emission line fluorescent X-ray energies and the normalized X-ray incident angle ($\alpha_{i} / \alpha_{c}$) for the SPM monolayer over an \ce{LaCl3} subphase (40 $\mu$M, pH 3.0) in the presence of 100 mM KCl. The absence of La emission lines (e.g., $L\alpha$ 4.6 keV, $L\beta$ 5.0 keV) across the spectrum, especially near $\alpha_{c}$, indicates the lack of surface enrichment of \ce{La^{3+}}. The characteristic emission lines of elements in the salt ingredients are labeled on the plots and only exhibit for $\alpha_{i} > \alpha_{c}$. These lines include chlorine $K\alpha$ ($\approx 2.6$ keV) and $K\beta$ ($\approx 2.8$ keV), and potassium $K\alpha$ ($\approx 3.3$ keV) and $K\beta$ ($\approx 3.6$ keV).
}
\label{fig:lacl3}
\end{figure}

The binding of Fe in the presence of co-solute salts to the neutral zwitterionic is unique to the PC template. Control experiments with a neutral monolayer formed by lipids (i.e., 1-hexadecanol) terminating with a alcohol group (R-OH) do not show Fe binding under similar conditions. NTXRF and X-ray reflectivity (XRR) measurements on a densely packed ($\Pi$ $\approx$ 40 mN/m) monolayer formed by 1-hexadecanol under identical subphase conditions ([\ce{FeCl3}] = 40 $\mu$M, [KCl] = 100 mM, pH 3.0) reveal no evidence of interfacial Fe binding to the alcoholic template (data not shown). Notably, iron-binding to other templates, such as those with carboxylic or phosphate groups, does not require co-solutes, as reported elsewhere \cite{Wang2011a,Wang2011b}. These findings align with other studies of iron binding to pH-controlled charge-neutral surfaces of fatty acid monolayers in the presence of KCl and Tris \cite{Wang2012a}. However, these experiments were conducted at significantly higher iron concentrations. The absence of surface signals from KCl is apparent. Previous NTXRF experiments under highly negatively charged phosphatidylinositol-4,5-bisphosphate monolayers have shown strong signals from potassium $K\alpha$ and $K\beta$ emission lines at the ML/solution interface \cite{Graber2015}. Similarly, experiments on positively charged 1,2-dipalmitoyl-3-trimethylammonium-propane monolayers show Cl presence at the interface \cite{Sung2015}. Therefore, the absence of low Z co-ions (e.g., \ce{K+}, \ce{Cl-}) at the SPM template is not due to the detection limits.

The binding of Fe to SPM induced by co-solutes is not yet fully understood. Our results show that co-ions do not bind to the interface but play a crucial role in the bulk, making the iron-specific binding to the zwitterionic PC template perplexing. To further examine Fe$^{3+}$ binding to a charged or neutral ML, we substituted LaCl$_3$ for \ce{FeCl3} in the subphase solutions (at 40 $\mu$M) under otherwise identical conditions (i.e., [KCl]=100 mM, pH 3.0, SPM-laden aqueous surfaces at $\Pi\approx35$ mN/m), conducting a control experiment. \cite{Wang2011b} Figure \ref{fig:lacl3} shows NTXRF data from such a monolayer, lacking La emission lines below the $\alpha_{c}$, evidence of no La binding to the charge neutral SPM at the interface. Previous experiments comparing Fe$^{3+}$ and La$^{3+}$ binding to carboxylic templates show that \ch{La^3+} behaves as a classical charged ion obeying Poisson-Boltzmann's equation, whereas Fe behaves differently \cite{Israelachvili2011, Bu2006}. These studies suggested that Fe$^{3+}$ in aqueous solutions bind to the carboxylic group through quasi-covalent bonds. This demonstrates that the binding of Fe(III) to the PC is not purely electrostatic in nature. 

\subsection*{pH effect on iron binding to SPM}
\begin{figure*}[hbt]
\centering
\includegraphics[width=0.9\textwidth]{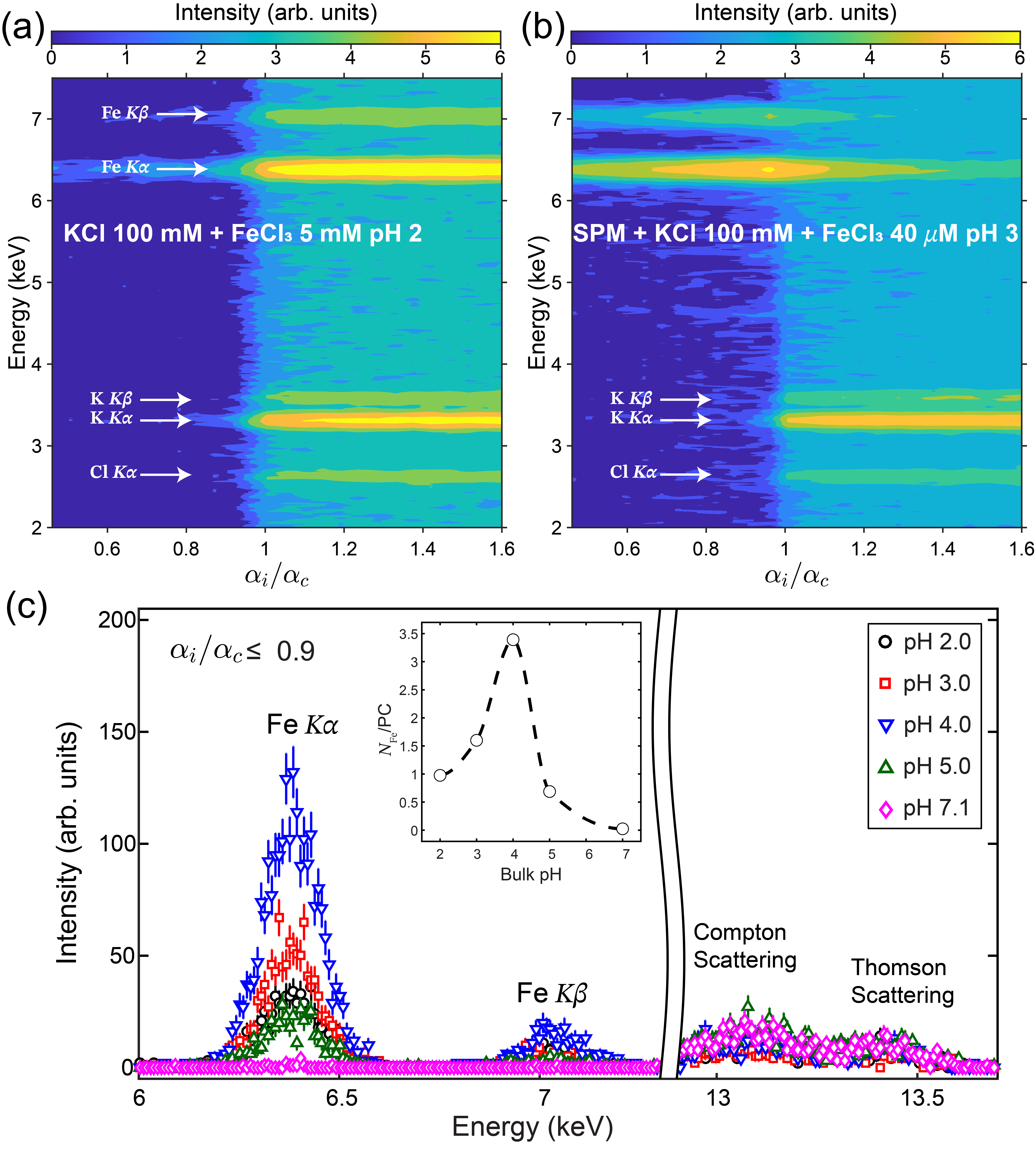} 
\caption{\textbf{Optimal Fe(III) binding to SPM occurs at pH 4.}
Two-dimensional fluorescence intensity contour plots as a function of emission line fluorescent X-ray energies and the normalized X-ray incident angle ($\alpha_{i} / \alpha_{ c}$). (a) A subphase (at pH 2) containing 100 mM KCl and 5 mM \ce{FeCl3} in the absence of SPM and (b) a subphase (at pH 3) containing 100 mM KCl and 40 $\mu$ M \ce{FeCl3} in the presence of an SPM monolayer showing strong Fe binding to SPM. (c) Surface fluorescence spectra for SPM over aqueous solutions (\ce{FeCl3} 40 $\mu$ M and KCl 100 mM) prepared at different pH levels (2.0, 3.0, 4.0, 5.0, and 7.1). The emission lines at 6.4 and 7.1 keV ($K\alpha$ and $K\beta$) are obtained by integrating the spectra over $\alpha_{i}$ less than $0.9\alpha_{c}$. The emission line intensities scale with Fe surface excess density ($\Gamma_{\rm Fe}$). The weak peaks labeled as Compton and Thomson scatterings result from the interaction of the incident beam with water and helium, showing that the incident beam is the same for all samples. The inset shows the number of Fe(III) ions per PC group ($N_{\rm Fe}$/PC) as a function of bulk pH, with optimum binding at pH 4.
}
\label{fig:fluores_APS}
\end{figure*}

Taking into account the above observations with solutions at pH 3, we further explore the influence of pH on Fe binding. Operation at a relatively low pH is essential due to the effect of pH on the species that Fe forms in aqueous solutions (iron-hydroxides).\cite{Wang2016} Previous studies have shown that the binding of iron-hydroxides to fatty acids or phospholipid monolayers is influenced by the bulk pH, particularly at low levels (i.e., pH 1-3) \cite{Wang2011a, Wang2011b}. Furthermore, iron binding to a DPPC monolayer, which shares the same PC head group as SPM, has been demonstrated to improve at pH 3.0. \cite{Wang2016}. 

To examine the effect of pH on Fe binding, we maintain constant salt concentrations ([\ce{FeCl3}] and [KCl]) and vary the solution pH from 2 to 7. Figure \ref{fig:fluores_APS} (a) shows an NTXRF color map of a relatively higher \ce{FeCl3} concentration (5 mM) with KCl (100 mM) in the absence of the SPM monolayer at pH 2. It displays signals from all ion species in the solution only above the critical angle ($\alpha_{c}$), confirming their presence in the bulk. Figure \ref{fig:fluores_APS} (b) presents a similar color map for a dilute solution of \ce{FeCl3} (40 $\mu$M) and KCl (100 mM) in the presence of an SPM template at the interface at pH 3. As shown above, Fe signals are prominent near the critical angle, revealing strong Fe binding under these conditions. Figure \ref{fig:fluores_APS} (c) shows the intensities versus the emission line energy (in the range of Fe $K\alpha$ ($\approx 6.40$ keV) and $K\beta$ ($\approx 7.06$ keV) obtained by integrating over $\alpha_{i}$ from 0 to $0.9\times \alpha_{c}$ at various pH levels, as indicated in the figure. The intensity of the emission lines is approximately proportional to the amount of Fe bound to the SPM under these conditions, with higher Fe binding leading to greater emission intensity. To quantify the iron surface adsorption at the interface, the data is analyzed by normalizing it to known iron concentrations in bulk solutions. The analysis yields Fe surface adsorption,  number of Fe(III) ions per PC group ($N_{\rm Fe}$/PC) as a function of bulk pH, as shown in the inset of Figure \ref{fig:fluores_APS}(c). Inspection of $N_{\rm Fe}$ shows maximum Fe binding at approximately pH 4, decreasing to negligible values (within uncertainty) at pH 7 (see Table \ref{table:Fe_count}). The panel to the right shows signals from the primary X-ray beam (at 13.3 keV) and Compton scattering from water, confirming that the incident X-ray flux at the surface for all conditions is the same within uncertainty.  

\begin{figure*}[ht]
\centering
\includegraphics[width=0.8\textwidth]{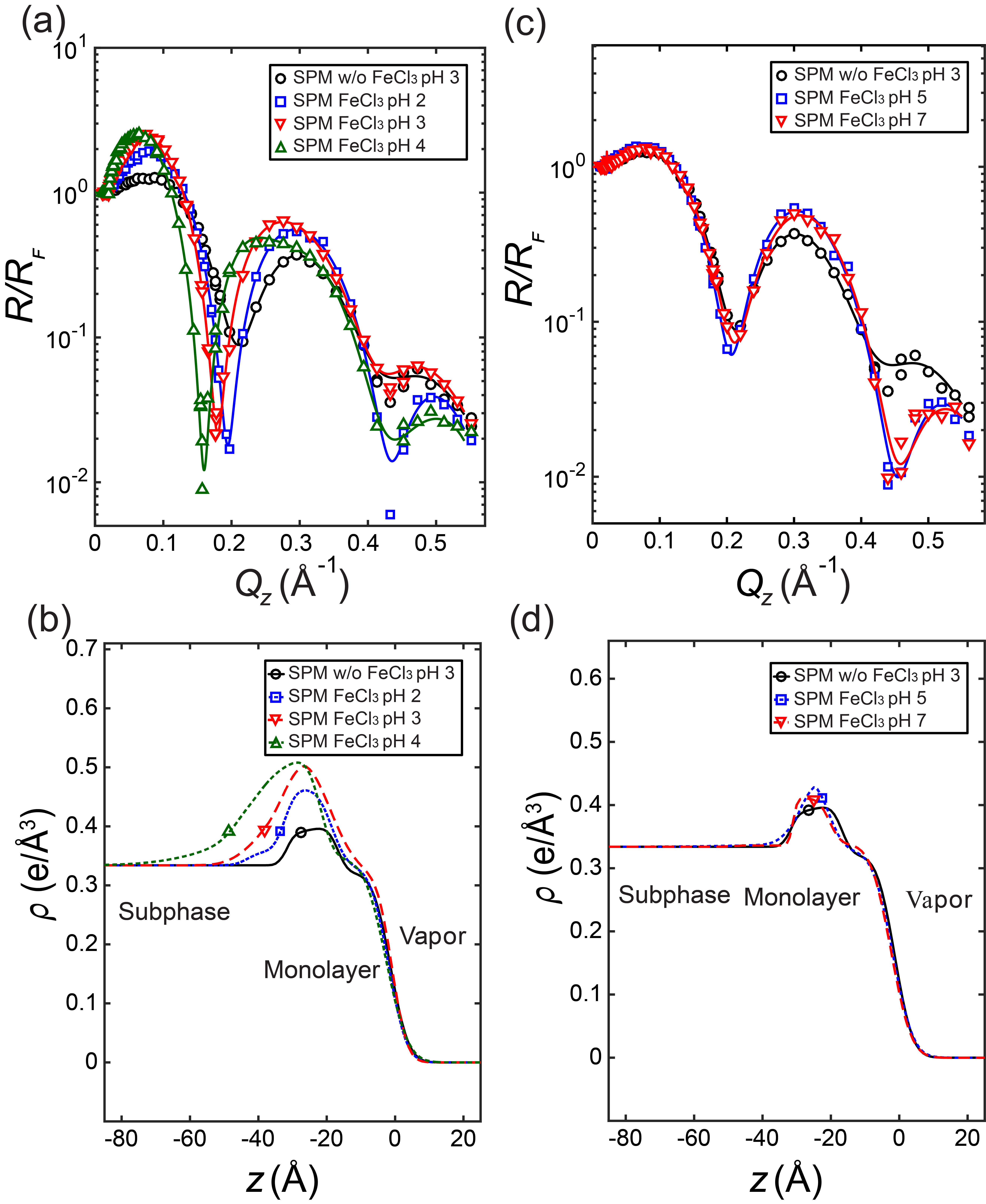}
\caption{\textbf{Confirming pH effect on Fe(III) binding by X-ray reflectivity.}
Normalized X-ray reflectivity ($R/R_{\rm F}$) for a monolayer of SPM over KCl (100 mM) subphase solutions in the presence and absence of \ce{FeCl3} at various pH levels. (a) $R/R_{\rm F}$ for SPM monolayers in a pure KCl subphase solution (100 mM) at pH 3 ($\bigcirc$) and KCl subphase solutions containing FeCl$_{3}$ (40 $\mu$M) at pH 2 ($\square$), 3 ($\bigtriangledown$), and 4 ($\bigtriangleup$). Solid lines are calculated $R/R_{\rm F}$ in terms of the ED profiles shown in (b). (b) ED profiles that best fit the $R/R_{\rm F}$ data in (a).  (c) $R/R_{\rm F}$ data for SPM monolayers in a pure KCl subphase solution (100 mM) at pH 3 ($\bigcirc$) and KCl subphase solutions containing FeCl$_{3}$ (40 $\mu$M) at pH 5 ($\square$) and 7 ($\bigtriangledown$).  (d) ED profiles that best-fit the $R/R_{\rm F}$ data in (c).}
\label{fig:APS}
\end{figure*}

\begin{figure*}[hbt]
\centering

\includegraphics[width=0.85\textwidth]{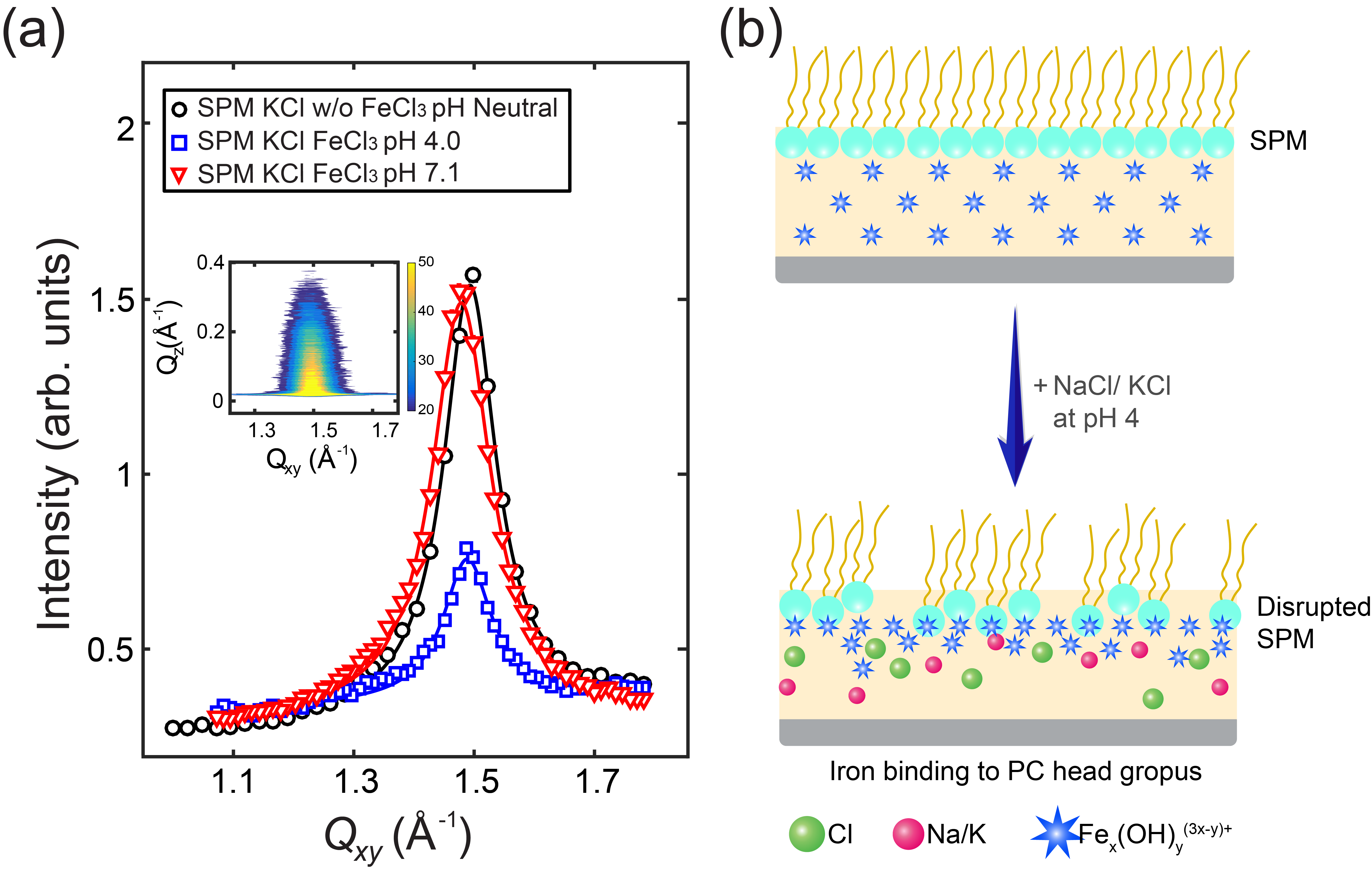}\\

\caption{\textbf{Fe (III) binding disrupts the structural integrity of the SPM monolayer.}
(a) Grazing incidence diffraction for SPM monolayer over KCl (100 mM) subphase solutions in the presence and absence of \ce{FeCl3} at various pH levels. The line-cut intensities, integrated over $Q_z$ below $0.2$ \text{\AA$^{-1}$}, are shown as a function of the in-plane wave vector transfer component $Q_{xy}$ for SPM over pure KCl (100 mM) subphase ($\bigcirc$), and in the presence of \ce{FeCl3} (40 $\mu$ M) at pH 7.1 ($\bigtriangledown$) and pH 4.0 ($\square$). The peak at about 1.5 \text{\AA$^{-1}$} is due to the hexagonal ordering of alkyl chains of SPM. Without Fe in the solutions and with Fe at pH 7, the diffraction peak is intense, while at pH 4, where Fe binding is strongest, the peak is much less intense. This suggests that the monolayer experiences deteriorated in-plane ordering upon Fe binding to SPM. The inset shows a two-dimensional GID intensity contour plot from SPM monolayers on an aqueous solution containing 100 mM KCl without pH treatment. (b) Illustrations of (top) SPM monolayer on KCl solution without Fe, and (bottom) Fe-hydroxide clusters bound to SPM in the presence of KCl.
}
\label{fig:APS_GIXD}
\end{figure*}

To further corroborate the NTXRF results, we conducted XRR measurements on the same films. Although XRR is not element-specific, it provides electron density (ED) profiles across the interface, which can be significantly altered by electron-rich, heavier-Z elements such as Fe. 

Figure \ref{fig:APS} (a) shows the normalized reflectivity, $R/R_{\rm F}$, for SPM films in the absence of \ce{FeCl3} (\(\bigcirc\)), and for the SPM films in the presence of \ce{FeCl3} at pH 2 (\(\square\)), pH 3 (\(\bigtriangledown\)), and pH 4 (\(\bigtriangleup\)). Without surface-bound Fe, the $R/R_{\rm F}$ profile arises mainly from the PC group of SPM, resulting in relatively low maxima. In contrast, the presence of surface-bound Fe significantly increases the $R/R_{\rm F}$ maxima. 

Quantitative analysis of the data using well-established methods \cite{Nielsen2011, Tolan1999, Vaknin2003a, Pershan2012} yields ED profiles across the interface. Table \ref{table:ref_params} summarizes the best ED profile parameters that fit the experimental XRR data using effective density model.\cite{Tolan1999} Figure \ref{fig:APS} (b) presents the ED profile of the monolayers extracted from fits to the $R/R_{\rm F}$ data in Figure \ref{fig:APS} (a) (solid lines through the experimental data). These profiles show a higher ED bell-shaped peak that can be associated with the PC head group and the Fe(III) complexes bound to it. Without Fe at pH 3, the analysis yields a relatively small bell-shaped peak merely due to the PC group. In the presence of Fe in the solution (including 100 mM KCl), there is a substantial increase in the ED in the head group region, confirming Fe binding to the PC group. The area under the bell-shaped ED provides an estimate of the amount of Fe bound to the ML. Notably, at pH 4, this area is the largest, presumably due to more Fe binding per PC head group. Additionally, the bell-shaped ED at pH 3 and 4 extend into the subphase much more significantly than the bare PC group with \ce{FeCl3}. These observations suggest that Fe binding to each head group likely involves a distribution of Fe-hydroxide clusters. This is consistent with previous results showing that Fe binding to charged interfaces results in bound Fe hydroxide clusters \cite{Wang2011a, Wang2011b,sowah2021iron}. The XRR results also confirm insignificant Fe binding at pH levels between 5 and 7. Figure \ref{fig:APS} (c) shows the $R/R_{\rm F}$ data for SPM monolayers with and without \ce{FeCl3} in solutions at pH 5 and 7. Qualitatively, these three reflectivity scans are nearly identical, indicating minimal  Fe at the interface. Further analysis of the $R/R_{\rm F}$ data yields the ED profiles shown in Figure \ref{fig:APS}(d). The bell-shaped peak in the ED profile corresponding to the head group remains nearly identical across all three films, confirming minimal Fe binding to the SPM monolayer at these pH levels. The analysis of the ED profiles, expressed in terms of the number of Fe(III) ions per PC group, is summarized in Table \ref{table:Fe_count}. It is important to note that, since XRR is not element-specific, the calculated Fe(III)/PC ratio depends on the specific iron species (FeO$_x$H$_y$) bound to the PC, as indicated in Table \ref{table:Fe_count} using the method provided in Ref.~[\!\!\citenum{Wang2012a}].

The massive Fe-cluster binding at pH 3 and 4 is expected to affect the SPM in-plane structure. Indeed, this is evident in grazing-incidence X-ray diffraction (GIXD) measurements of the film before and after Fe binding. Figure \ref{fig:APS_GIXD} (a) shows the diffraction pattern at a grazing angle of incidence of the bare monolayer, consisting of a single strong peak commonly associated with the hexagonal ordering of hydrocarbon chains of the SPM. As Fe binds to the SPM at pH 4, this Bragg reflection is significantly diminished, indicating an in-plane disruption of the SPM monolayer. We also note that the diffraction peak at pH 7 shows a subtle variation compared to the pure SPM monolayer without Fe binding, aligning with the observations mentioned above.\\
\section{Conclusions}
Using surface-sensitive X-ray diffraction and spectroscopic measurements, we have determined the conditions under which iron ions (Fe(III)) bind to a biologically derived SPM membrane. This membrane, composed of SPM extracted from bovine spinal cord, was formed as a Langmuir monolayer at the aqueous-vapor interface. Our findings demonstrate that iron ions, or their complexes (such as iron-hydroxides), bind effectively to the PC template of SPM only in the presence of co-solutes, such as NaCl or KCl, at physiological concentrations ($\sim$100 mM). In the absence of these salts, iron ions show little to no affinity for the PC template. We also found that the accumulation of iron correlates with salt concentration, reaching saturation at physiological levels. Furthermore, pH plays a critical role in iron binding, with maximum binding efficiency occurring around pH 4. The pH-dependent behavior of lipid monolayers, especially at low pH, is highly relevant to various pathological conditions, as numerous biological processes, particularly in membranes, are sensitive to local pH fluctuations. Interestingly, other multivalent ions, such as La$^{3+}$ or Ca$^{2+}$, do not accumulate in the PC template under similar conditions, instead behaving according to classical Poisson-Boltzmann theory in solution. This stark difference in behavior suggests that the mechanism of iron binding is unique. We propose that iron binds through the formation of Fe-hydroxides, likely involving quasi-covalent interactions rather than purely electrostatic binding. The exclusive binding of iron ions to the SPM template in the presence of additional salts is intriguing and warrants further investigation to uncover the underlying mechanisms. Moreover, iron-binding also affects the organization of SPM molecules in the membrane, causing disruption of the SPM membrane. Although our model system represents an idealized SPM monolayer, a component of the myelin sheath rather than a complete myelin membrane, we propose that iron binding to SPM may influence membrane integrity and could potentially contribute to myelin sheath damage in neurological disorders. However, this hypothesis should be approached with caution due to the simplified nature of our ideal experimental model.\\
\section{Methods}

\noindent{\it\bf Reagents and Materials}
Sphingomyelin (C$_{41}$H$_{83}$N$_{2}$O$_{6}$P, CAS\# 85187-10-6, natural source: bovine spinal cord) was purchased from Matreya LLC and was dissolved in a 3:1 chloroform/methanol stock solution.
All aqueous solutions were prepared with ultrapure water (Millipore, Milli-Q, and NANOpure, Barnstead; resistivity 18.1 M$\Omega$ cm, pH $\sim$5.6).  Hydrochloric acid (HCl) and Potassium hydroxide (KOH) were used to vary the pH between 2 and 7. All solutions were kept at constant temperature (20$^{\circ}$C) in the Langmuir trough during all measurements (i.e., isotherms and X-ray characterizations). KCl (CAS\# 7447-40-7) and FeCl$_3\cdot$6H$_2$O (CAS\# 10025-77-1), NaCl, KI, CalCl$_2$ and LaCl$_3$, were purchased from Fisher Scientific and used as obtained. In some cases, concentrated stock solutions were used to adjust the concentration of samples.

\noindent{\it\bf Sample Environment}
Aqueous solutions for surface characterizations were contained in a Langmuir trough contained in an air-tight chamber. The chamber is purged with water-saturated helium to be exchanged with air at an oxygen level lower than 1\% in atmospheric conditions. The oxygen sensor S101, Qubit System Inc., was used to monitor oxygen levels.

Surface pressure ($\Pi$) versus molecular area ($A_{\rm mol}$) isotherms were performed for SPM spread over the aqueous surfaces prior to any surface X-ray measurements. The SPM monolayer was spread at large molecular areas (in the 2D gas phase) and subsequently compressed with a motorized Teflon barrier. Surface pressure was measured with a filter-paper Wilhelmy plate and maintained at $\sim 30$ mM/m (corresponding molecular area  $43 \pm 2$ {\AA$^2$} obtained from pressure-area isotherms.) for all X-ray characterizations (both synchrotron-based and in-house facility-based) with a predetermined tolerance of $\pm 3$ mN/m. The profiles of the isotherms are all similar and resemble those reported in previous studies.\cite{Vaknin2001a}

In our experiments, two distinct protocols were employed to add co-solutes to the \ce{FeCl3} subphase with the SPM monolayer. The first protocol involves the addition of co-solute salts directly to an existing \ce{FeCl3} subphase with a monolayer. This method requires a longer waiting time to allow for the diffusion of the co-solute salt to the area of the X-ray footprint because the salt needs to traverse the barrier before it can interact with the monolayer. The second protocol, designed to save time during pH-dependent measurements, involves the initial preparation of a solution containing both \ch{FeCl3} and the desired co-solute salt. This premixed solution is then used to spread the monolayer. By starting with the \ch{FeCl3} and co-solute salt already combined, this approach eliminates the waiting period associated with the diffusion process, thereby expediting the overall experimental timeline. We have verified that both protocols yield the same results, ensuring the consistency and reliability of our experimental data regardless of the protocol employed.

\noindent{\it\bf Surface-sensitive X-ray Measurements}

Surface-sensitive synchrotron X-ray diffraction and spectroscopic measurements, i.e., X-ray reflectivity (XRR), X-ray near total reflection X-ray fluorescence (NTXRF) and grazing incidence X-ray diffraction (GIXD), were conducted on the liquid surface spectrometer (LSS) located at sector 9ID-C, Advanced Photon Source (APS), Argonne National Laboratory. The highly collimated and monochromatic X-ray beam (wave vector denoted as $\bf{k_{\rm i}}$, photon energy $E$=13.474 keV and wavelength $\lambda=0.9201$ {\AA}) was incident on the surface ($x$-$y$ coordinates) of aqueous solutions at an incident angle $\alpha_{i}$.  A point detector (Bicron scintillation detector) was used to collect the scattered X-ray waves (wave vector denoted as $\bf{k_{\rm f}}$ ) from the aqueous surface at an exit angle $\alpha_{f}$.  For specular x-ray reflectivity measurements in which $\alpha_{i}=\alpha_{f}$, the transfer of the X-ray wave vector, $\bf {Q}$, is $\bf Q= k_{f}-k_{i}$, is expressed in our coordinate system as $ {\bf Q}=(0,0, Q_z)$, where $Q_z$ is the $z$-component along the surface normal ($z$-axis) and $Q_z=(4\pi/\lambda)\sin\alpha_{i}$. The XRR results are displayed as $R (Q_{z})/R_{\rm F}$, where, $R_{\rm F}$ is the Fresnel reflectivity that is calculated for an ideally smooth and sharp aqueous/vapor interface.  The electron density (ED) profiles, $\rho$, across the interfaces are extracted from fits to the $R/R_{\rm F}$ data using Parratt's recursive method\cite{Nielsen2011}.  The effective density model \cite{Tolan1999} is used to construct parametric ED profiles for structural refinement (see examples elsewhere\cite{Wang2014}).

GIXD was measured at constant incident angle $\alpha_{i}=0.065^{\circ}$ that is below the critical angle $\alpha_{c} \simeq 0.09^{\circ}$ and with a position sensitive detector (PSD, Mythen) placed at the in-plane diffraction angle, $2\theta$, collimated with Soller slits. The GIXD intensities are recorded as a function of the modulus of the in-plane scattering vector component, $Q_{xy}=(4\pi/\lambda)\sin\theta$ and the vertical component $Q_z$.

For NTXRF measurements, an energy dispersive detector (EDD, Vortex-90EX) was placed above the aqueous surfaces with the detecting probe subtending the illuminated area of the X-rays.\cite{Vaknin2003a,Wang2011b}
A collimator was placed in front of the probe to collect the X-rays (fluorescent or scattered X-rays) along the normal direction of the surface (i.e., $\alpha_{f}\approx90^{\circ}$) with an acceptance angle $\sim 1^{\circ}$. NTXRF method is both element- and surface-sensitive.\cite{Vaknin2003a, Bu2009, Wang2011a, Wang2011b, Wang2014}   
In NTXRF measurements, the incident angle $\alpha_{i}$ is varied across the critical angle for total reflection, $\alpha_{c}$.  When $\alpha_{i}<\alpha_{c}$, the X-rays only penetrate a shallow depth in water ($\sim 50$ {\AA} as $\alpha_{i}\rightarrow 0$) thus probing the species mainly at the interface.  For $\alpha_{i}>\alpha_{c}$, the X-rays penetrate into the bulk water to a few microns, resulting in fluorescence signals from the bulk.
By comparing the NTXRF spectra from bare-surface \ch{FeCl3} solution of known concentrations to that with a SPM ML where Fe(III) aggregates in the immediate vicinity of ML, the surface adsorption (i.e., surface excess) of Fe, $\Gamma_{\rm Fe}$, can be determined.\cite{Wang2011b}

In this study, typical synchrotron X-ray measurements take $\sim 20$ min for a reflectivity measurement ($Q_z$ up to 0.55 {\AA}$^{-1}$), $\lesssim 10$ mins for a fluorescence measurement ($15$ sec at each $\alpha_{i}$) and $\sim 20$ min for a GIXD measurement; therefore, they are considered a relative fast probe for surface structure formed after spreading and compression of ML. 
X-ray reflectivity and fluorescence spectroscopy measurements were also performed with an in-house LSS at Ames National Laboratory using a rotating anode source (Cu-target, incident X-ray energy $E=8.048$ keV).\cite{Vaknin2003a, Pershan2012}

\section{Acknowledgements}
We thank Dr. Ivan Kuzmenko at beamline 9ID-C, APS, Argonne National Laboratory, for beamline technical support. We also thank Noah Hagen and Alex De Palma at Ames National Laboratory for help with the initial preliminary experiments, and A.D.P. was supported by the Science Undergraduate Laboratory Internship (SULI) program of the U.S. Department of Energy.  Research in Ames National Laboratory is funded by the U.S. Department of Energy, Office of Basic Energy Sciences, Division of Materials Sciences and Engineering.  Ames National Laboratory is operated for the U.S. Department of Energy by Iowa State University under Contract No. DE-AC02-07CH11358. Use of the Advanced Photon Source, an Office of Science User Facility operated for the U.S. Department of Energy (DOE) Office of Science by Argonne National Laboratory, was supported by the U.S. DOE under Contract No. DE-AC02-06CH11357.

\section{Author contributions statement}
D.V. initiated the project and, together with W.W., designed the experiments. H.Z., D.V., and W.W. conducted the experiments. Data analysis was performed by D.V., B.N., and W.W.. D.V., B.N., and W.W. wrote the manuscript. All authors read and reviewed the manuscript.
\section{Additional information}
The corresponding author declares that they have no competing financial interests on behalf of all authors of the paper. 

\section{Supporting Information}
The Supporting Information is available free of charge on the Elsevier's website at DOI: xxxxx/yyyyy

\section{Data Availability}
The findings of this study are supported by data found within the article and the provided Supplementary Information. Further relevant information and source data can be obtained from the following DOI: https://doi.org/10.7910/DVN/PD8UVT

\bibliography{SPM.bib}

\clearpage
\onecolumn

\setcounter{page}{1}
\setcounter{figure}{0}
\setcounter{equation}{0}
\setcounter{table}{0}

\renewcommand{\thefigure}{S\arabic{figure}}
\renewcommand{\theequation}{S\arabic{equation}}
\renewcommand{\thetable}{S\arabic{table}}
\renewcommand{\thepage}{S\arabic{page}} 

\makeatletter
\renewcommand\@biblabel[1]{[S#1]}
\makeatother

\section{Supporting information}

\begin{center} 
	{\textbf{\LARGE{Specific Iron Binding to Natural Sphingomyelin Membrane Induced by Non-Specific Co-Solutes}}}\\
	\bigskip
	\normalsize
    Wenjie Wang,$^{\dagger}$
    Honghu Zhang,$^\ddagger$
    Binay P. Nayak,$^\mathparagraph$
	David Vaknin,$^{\ast ,\S}$\\
 
	\bigskip

	{$\dagger$\it Division of Materials Sciences and Engineering, Ames National Laboratory, Ames, Iowa 50011, United States}\\
	{$\ddagger$\it  Ames National Laboratory, and Department of Materials Science and Engineering, Iowa State University, Ames, Iowa 50011, United States}\\
	{$\mathparagraph$\it  Ames National Laboratory, and Department of Chemical and Biological Engineering, Iowa State University, Ames, Iowa 50011, United States}\\
	{$\S$\it  Ames National Laboratory, and Department of Physics and Astronomy, Iowa State University, Ames, Iowa 50011, United States}\\
	\bigskip
	{E-mail: vaknin@ameslab.gov}\\
\end{center}

\subsection{Additional details on X-ray experiments}

\begin{figure*}[hbt]
\centering

\includegraphics[width=0.95\textwidth]{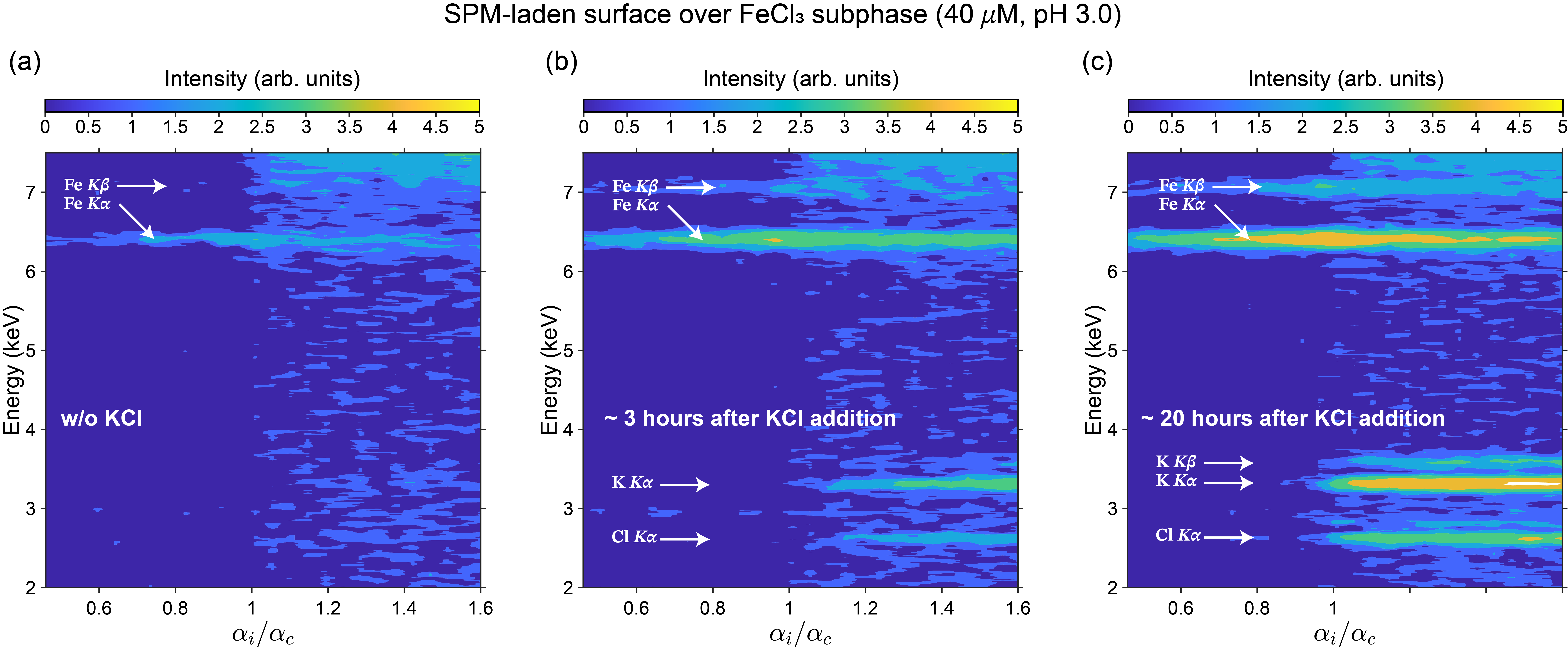}\\

\caption{
Two-dimensional fluorescence intensity contour plots as a function of emission line fluorescent X-ray energies and the normalized X-ray incident angle ($\alpha_{i} / \alpha_{ c}$), for SPM-laden surface over \ce{FeCl3} subphase ($40$ $\mu$M, pH 3.0). The subphase conditions, in addition to containing \ce{FeCl3}, are (a) without other salts; (b) injected KCl ($\sim 45$ mM) in diffusion for $\sim$ 3--5 hrs ; (c)  KCl diffusion stabilized after $\sim 20$ hrs. Panel (a) and (c) replicate Figure \ref{fig:fluores_Ames_2D} and are juxtaposed with panel (b)  for visual comparison.}
\label{fig:KCl_compare}
\end{figure*}

To examine the impact of KCl on Fe adsorption, KCl was injected into the subphase beneath the monolayer in a Langmuir trough, ensuring it did not directly intersect the X-ray beam. Multiple NTXRF experiments were conducted consecutively to track Fe adsorption as KCl diffused into the irradiated area. Figure \ref{fig:KCl_compare} depicts the temporal progression of Fe binding, reflecting the diffusion of KCl. Moreover, the reverse protocol, where the monolayer was spread onto Fe and KCl solutions directly, produced comparable final Fe binding results and significantly shortened subsequent experiments involving other salts and pH variations. Therefore, this reverse protocol was adopted to investigate the effects of pH and various co-solute salts. 

\begin{figure*}[hbt]
\centering
\includegraphics[width=0.95\textwidth]{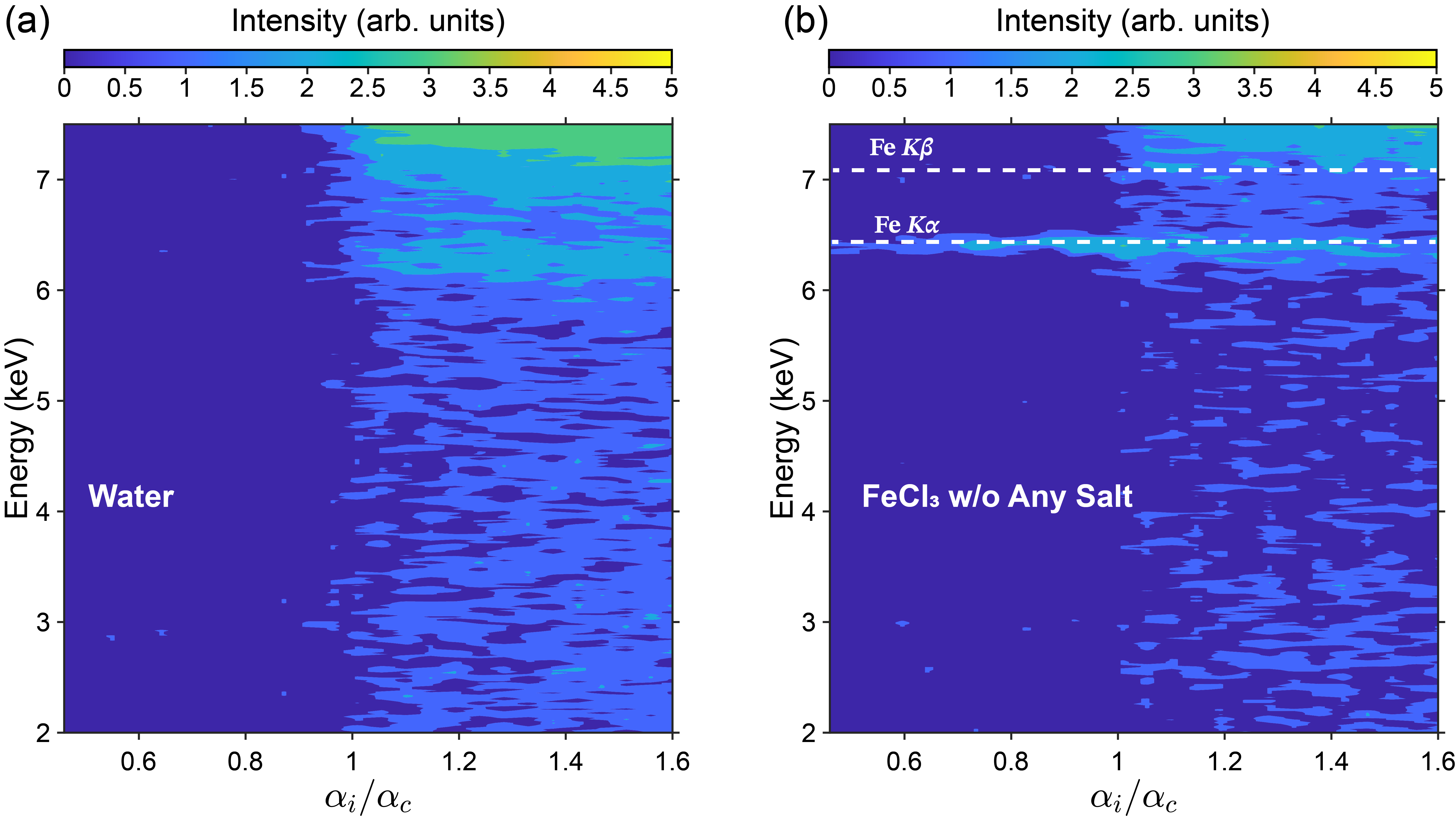}\\
\caption{
Two-dimensional fluorescence intensity contour plots as a function of emission line fluorescent X-ray energies and the normalized X-ray incident angle ($\alpha_{i} / \alpha_{c}$), for (a) pure water (b) SPM-laden surface over \ce{FeCl3} subphase ($40$ $\mu$M, pH 3.0). The intensity around $\sim 6.3$ keV can be partly attributed to the escape peak typically observed in Si-based energy dispersive detectors due to the 8.05 keV incident X-rays. Panel (b) replicates Figure \ref{fig:fluores_Ames_2D}(a) and is juxtaposed with panel (a) to facilitate visual comparison.}
\label{fig:water}
\end{figure*}

In the absence of co-solutes such as KCl, the adsorption of Fe onto the SPM monolayer from dilute \ch{FeCl3} solutions is negligible. As shown in Figure \ref{fig:water} (a), the spectrum of pure water exhibits no emission lines. When SPM is spread over a dilute \ch{FeCl3} solution, the spectra such as in Figure \ref{fig:water} (b) remain virtually identical to that of pure water, indicating minimal Fe adsorption. This stands in stark contrast to the spectrum of SPM over \ch{FeCl3} solutions containing KCl, shown in Figure \ref{fig:fluores_Ames_2D} (b), where distinct emission lines indicate that KCl significantly enhances Fe adsorption onto the SPM surface.

\subsection{Additional details on XRR analysis}

\begin{figure*}[hbt]
\centering
\includegraphics[width=0.95\textwidth]{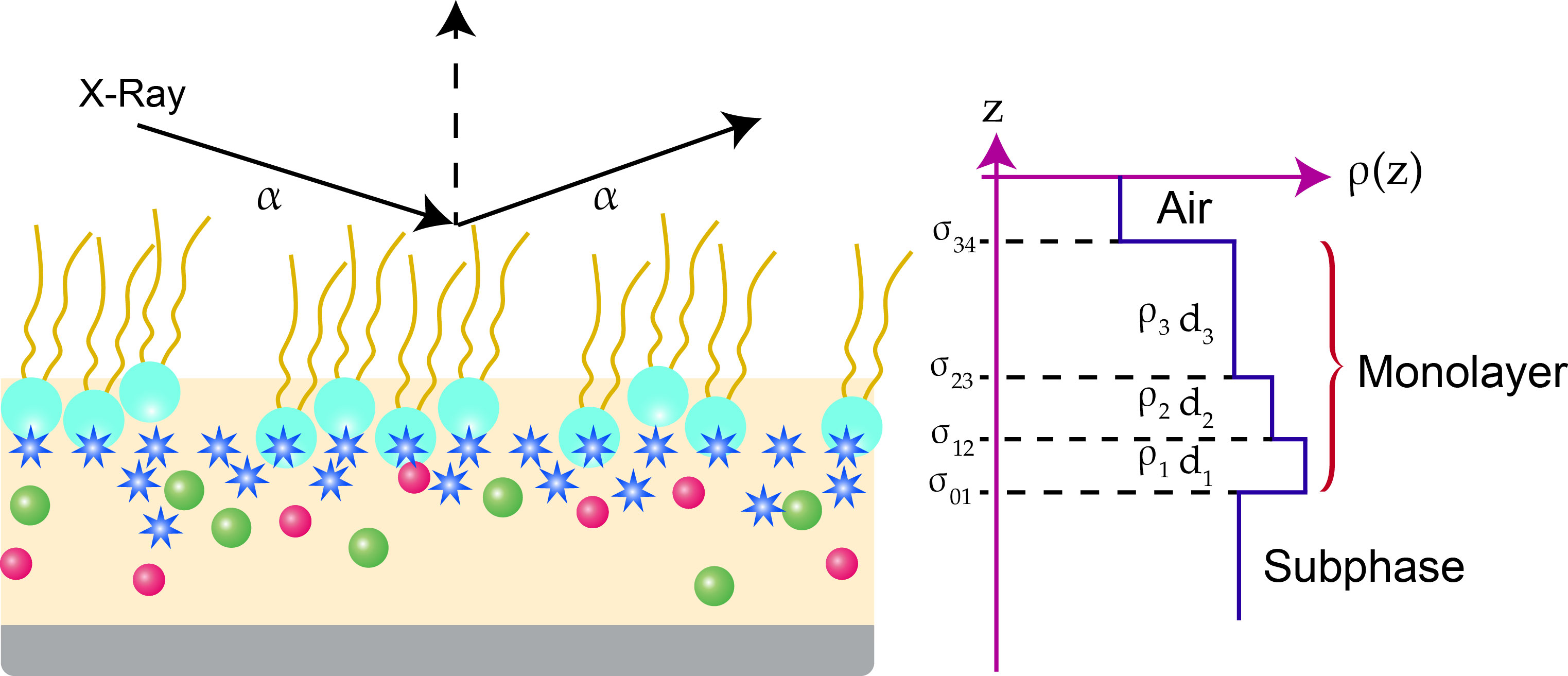}\\
\caption{Schematic representation of X-ray reflectivity (left) and electron density profile (right) normal to the surface for an SPM monolayer over the liquid subphase. The electron density profile is approximated with an effective density model, a common approach in X-ray reflectivity analysis.\cite{Tolan1999} The model starts with a discrete profile (i.e., $\sigma_{ij}\rightarrow 0$ for sharp interfaces), but after data refinement, the resulting ED profile is typically smooth and continuous, allowing for more accurate representation of the interfacial structure. }
\label{fig:box_model}
\end{figure*}

X-ray reflectivity (XRR) data were analyzed using an effective density model \cite{Tolan1999}. This model represents the system as a series of stacked slabs above a subphase, as shown in Figure \ref{fig:box_model}. Each slab is defined by its thickness ($d_i$) and uniform ED ($\rho_i$), while the interfacial roughness between adjacent slabs is described by the parameter $\sigma_{ij}$. The complexity of the monolayer system being studied dictates the total number of slabs required for the model.

The effective density model, characterized by parameters such as $\rho_i$, $d_i$, and $\sigma_{ij}$, offers a flexible method for constructing an ED profile to fit the experimental reflectivity data $R(Q_z)$. Unlike the box model \cite{Vaknin2003a}, which imposes stricter constraints on these parameters, the effective density model allows greater freedom in generating ED profiles. For instance, there is no requirement that $\sigma_{ij}$ must be significantly smaller than the thicknesses of adjacent $i$-th and $j$-th slabs, where $j=i\pm 1$. However, this flexibility comes with a trade-off in terms of interpretability. In the box model \cite{Vaknin2003a}, the parameters are grouped to represent physically meaningful structures, such as head or tail groups, with constraints based on known dimensions and compositions. Despite the higher degree of correlation between parameters in the effective density model, the resulting ED profiles still converge to a consistent best-fit solution, showing minimal variability.

Table \ref{table:ref_params} lists the specific parameters of the effective density model used to generate the ED profiles displayed in Figure \ref{fig:box_model}, which best match the experimental $R(Q_z)$ data shown in Figure \ref{fig:APS}.

\begin{table*}[htbp]
\caption{Parameters for the effective density model that generates the ED profiles shown in Figure \ref{fig:box_model} and calculated $R(Q_z)/R_{\rm F}$ that match the experimental data shown in Figure\ref{fig:APS} (c) and (d).} 
\begin{threeparttable}
\scriptsize
\centering
    \begin{tabular}{lc|cccccccccc}
    \toprule

    Surface Condition & pH & $\sigma_{01}$ & $d_1$ & $\rho_1$ & $\sigma_{12}$ & $d_2$ & $\rho_2$ & $\sigma_{23}$ & $d_3$ & $\rho_3$ & $\sigma_{34}$ \\
    \midrule
    SPM w/o \ch{FeCl3}$^*$ & 3 & 1.7 & 15.2 & 0.399 & 2.3 & 15.5 & 0.321 & 4.1& $--$& $--$& $--$ \\
    \midrule
    \multirow{5}{*}{SPM + \ch{FeCl3}} & 2 & 5.6 & 11.3 & 0.390 & 5.3 & 12.1 & 0.559 & 6.7 & 18.7 & 0.317 & 3.5 \\
    & 3 & 3.0 & 17.9 & 0.336 & 1.0 & 10.7 & 0.411 & 2.5 & 18.0 & 0.338 & 4.7 \\
    & 4 & 16.6 & 3.0 & 0.430 & 5.2 & 20.0 & 0.540 & 3.2 & 19.4 & 0.345 & 5.0 \\
    & 5 & 3.0 & 10.7 & 0.357 & 2.6 & 13.9 & 0.464 & 3.5 & 16.5 & 0.331 & 4.2 \\
    & 7 & 16.5 & 14.0 & 0.339 & 4.6 & 6.1 & 0.506 & 3.9 & 19.9 & 0.327 & 4.6 \\
    \bottomrule
    \end{tabular}
\begin{tablenotes}
    \item[*] {\scriptsize Fewer parameters are necessary to fit the $R(Q_z)$ data sufficiently well for pure SPM monolayers on pure water.}
\end{tablenotes}
\end{threeparttable}
\label{table:ref_params}
\end{table*}

\subsection{Calculation of Number of Fe(III) per PC group}
\begin{table}[htbp]
\caption{Comparison of number of iron ions per SPM based on XRR and NTXRF analysis. The error analysis for the XRR analysis was conducted using Monte Carlo methods in a parametric space. }
\label{table:Fe_count}
\begin{threeparttable}
\scriptsize
\centering
\begin{tabular}{@{}|c|c|c|c|c|@{}}
\toprule
SPM\tnote{a} + \ch{FeCl3} (pH) & $\Gamma_{\textrm e}$ [e/\AA$^2$]) & $N_{\textrm Fe}^{\rm NTXRF} $ & $N_{\rm Fe(OH_2)_6^{3+}}^{\rm XRR} $ & $N_{\rm Fe(OH)_3}^{\rm XRR} $ \\
\midrule
  2 &   2.2 ± 0.1 &       1.0 ± 0.2 &             0.7 ± 0.1 &            1.6 ± 0.2 \\
  3 &   3.4 ± 0.2 &       1.6 ± 0.2 &             1.5 ± 0.2 &            3.1 ± 0.4 \\
  4 &   4.2 ± 0.1 &       3.4 ± 0.5 &             1.9 ± 0.1 &            4.0 ± 0.3 \\
  5 &   1.0 ± 0.1 &       0.7 ± 0.1 &             \textless 0.1 &        \textless 0.1 \\
  7\tnote{b}  &   0.8 ± 0.1 &   \textless 0.1 &         N/A &                   N/A \\
\bottomrule
\end{tabular}
\begin{tablenotes}
    \item[a] {\scriptsize $\Gamma_{\textrm e}$ for pure SPM monolayer is 0.9 ± 0.2 e/{\AA}$^2$.}
    \item[b] {\scriptsize $\Gamma_{\textrm e}$ for pure SPM + \ch{FeCl3} at pH 7 is too close to that of pure SPM within uncertainty to derive the number of iron ions per SPM.}
\end{tablenotes}
\end{threeparttable}
\end{table}

The ED profiles obtained from XRR are inherently non-element-specific, which limits our ability to directly quantify iron adsorption using XRR alone. Therefore, quantifying the iron adsorption becomes model-dependent, as the assignment of excess electrons in the ED profile can vary significantly depending on whether they are attributed to pure iron atoms or iron hydroxide species. In this study, we use two iron hydroxides, i.e., \ch{[Fe(H2O)6]^{3+}} and \ch{Fe(OH)3}, as examples to estimate the number of iron ions per SPM molecule, denoted as $N_{\rm Fe(O_xH_y)_n}^{\rm XRR}$, and compare these values with those obtained from NTXRF analysis, $N_{\rm Fe}^{\rm NTXRF}$.

We use Eq. (\ref{eq:gamma}) to estimate the density of surface excess  electrons in the monolayer with respect to the subphase, denoted as $\Gamma_{\rm e}$.\cite{Wang2012a} Iron adsorption per SPM can be estimated by calculating the difference in $\Gamma_{\rm e}$ between pure SPM and SPM with iron adsorbed, using Eq. (\ref{eq:Nfe}).

\begin{equation} 
    \Gamma_{\rm e} = \int \left[ \rho(z) - \rho_{\text{sub}}(z) \right] {\rm d}z
    \label{eq:gamma}
\end{equation}

\begin{equation} 
    N_{\mathrm{Fe(O_xH_y)_n}}^{\rm XRR} = \frac{A_{\rm mol} \times \left(\Gamma_{\rm e}^{\rm SPM+Fe(O_xH_y)_n} - \Gamma_{\rm e}^{\rm SPM}\right)}{n_{\mathrm{Fe(O_xH_y)_n}}^{\rm e} - \rho_{\mathrm{sub}} \times V_{\rm Fe(O_xH_y)_n}}
    \label{eq:Nfe}
\end{equation}

Here, $A_{\rm mol}$ is the molecular surface area of SPM, $V_{\rm Fe(O_xH_y)_n}$ is the volume per iron hydroxide replacing the same volume of the subphase, $\rho_{\rm sub}$ is the electron density of the subphase, and $n_{\mathrm{Fe(O_xH_y)_n}}^{\rm e}$ is the number of electrons per iron hydroxide. The volume of \ch{[Fe(H2O)6]^{3+}}, an octahedral complex, is estimated using the \ch{Fe-O} bond distance, $r$, and the corresponding volume of a sphere, $4\pi r^3/3$, with $r\approx 2$ {\AA}. The volume for \ch{Fe(OH)3} is estimated from its solid-state structure.\cite{Wang2012a} The results are summarized in Table \ref{table:Fe_count} for comparison with the NTXRF measurements.

\begin{center}
-End of Document-
\end{center}

\end{document}